\documentclass[12pt]{article}
\usepackage{amsfonts,amsthm,amsmath,amssymb,upgreek,bm}
\usepackage{dsfont}
\usepackage[paper=letterpaper,margin=.85in]{geometry}
\usepackage{graphicx}
\usepackage{units}
\usepackage{float}
\usepackage[export]{adjustbox}
\usepackage{bbold}
\usepackage[shortlabels]{enumitem}

\usepackage{color}
\usepackage[dvipsnames]{xcolor}
\definecolor{darkblue}{rgb}{0.1,0.1,.7}
\definecolor{purple}{rgb}{0.6,0,0.6}
\definecolor{orange}{rgb}{0.9,0.6,0}
\usepackage[colorlinks, linkcolor=darkblue, citecolor=darkblue, urlcolor=darkblue, linktocpage]{hyperref} 
\usepackage[square, comma, compress,numbers]{natbib}
\usepackage{physics}
\usepackage{tensor}
\usepackage{psfrag}
\usepackage{footmisc}
\usepackage{url}
\usepackage{mathtools}
\usepackage{tikz}
    \usepackage{amssymb,amsfonts,amsmath}
    \usepackage{tkz-euclide}
\usetikzlibrary{arrows.meta,calc,patterns}
\usepackage{pgfplots}
\usepackage[]{amsmath}
\usepackage[]{graphicx}
\usepackage[]{latexsym}
\usepackage[utf8]{inputenc}
\usepackage{geometry}
\usepackage{amscd}
\usepackage[all,cmtip]{xy}
\usepackage{mathrsfs}
\usepackage[most]{tcolorbox}
\usepackage{tikz}
\usepackage{subcaption}

\numberwithin{equation}{section}
\newcommand{\be}{\begin{equation}}
\newcommand{\ee}{\end{equation}}

\usetikzlibrary{arrows,positioning}

\def\SL2{\widetilde{SL}(2,\mathbb R)}

\numberwithin{equation}{section}

\newcommand {\bes} {\begin {equation*}}
\newcommand {\ees} {\end {equation*}}
\newcommand {\beq} {\begin {equation}}
\newcommand {\eeq} {\end {equation}}
\newcommand {\bea} {\begin {eqnarray}}
\newcommand {\ea} {\end {eqnarray}}
\newcommand {\eea} {\end {eqnarray}}

\numberwithin{equation}{section}

\renewcommand{\i}{{\rm i}}

\def\<{\langle}
\def\>{\rangle}

\tikzset{
    >=stealth',
    punkt/.style={
           rectangle,
           rounded corners,
           draw=black, very thick,
           text width=15em,
           minimum height=2em,
           text centered},
    pil/.style={
           ->,
           thick,
           shorten <=2pt,
           shorten >=2pt,}
}

 \def\ie{\begin{equation}\begin{aligned}}
\def\fe{\end{aligned}\end{equation}}

\renewcommand{\i}{\mathrm{i}}
\renewcommand{\d}{\mathrm{d}}

\begin{document}

\thispagestyle{empty}
\vspace*{2.5cm}
\begin{center}

{\bf{\LARGE Wormholes do not contribute to \\ \vspace{15pt} supersymmetric partition functions}}



\begin{center}
\vspace{1cm}
 {\bf Jan Boruch and Douglas Stanford}\\
\bigskip \rm
\bigskip
Leinweber Institute for Theoretical Physics at Stanford, Stanford, CA 94305, USA
\rm
\end{center}
\vspace{2.5cm}
{\bf Abstract}
\end{center}

\begin{quotation}
\noindent

We show that wormholes connecting multiple asymptotic boundaries always spontaneously break at least some of the symmetries preserved by their boundary conditions. Their configurations then form nontrivial orbits of the broken symmetry, and for the case of a fermionic symmetry, the contribution of each orbit vanishes. We conclude that wormholes do not contribute to moments of supersymmetric partition functions. This extends results on supersymmetric indices by Iliesiu, Kologlu, and Turiaci. We emphasize that supersymmetric partition functions can still have a nonzero ensemble variance. In SYK-like examples that we analyze, this arises due to the contribution of ``non-geometric wormhole'' configurations that preserve all continuous symmetries.
\end{quotation}

\setcounter{page}{0}
\setcounter{tocdepth}{2}
\setcounter{footnote}{0}
\newpage

\tableofcontents

\newpage

\section{Introduction}

It is not clear what role spacetime wormholes play in traditional examples of holography, based on specific boundary theories. An interesting recent proposal is that they could be related to a subtlety in the large $N$ limit of the boundary theory. In particular, it was proposed that they capture averaged properties of terms that depend on $N$ in an erratic way \cite{Schlenker:2022dyo,Liu:2025ikq, Kudler-Flam:2026nzz,Liu:2026fnd}. To test this idea, one would ideally find a boundary partition function or observable that can be computed exactly at finite $N$, identify the erratic part, and find a wormhole that contributes to the corresponding bulk calculation. To find the simplest setting for this test, one is led to wonder whether products of supersymmetric partition functions can get connected contributions from wormholes. 

This question was already answered (``no''), for the case of supersymmetric indices, by Iliesiu, Kologlu and Turiaci \cite{Iliesiu:2021are}. They showed that the Schwarzian theory capturing gravitational fluctuations in the approximate AdS$_2$ region associated to the index computation contains a fermionic zero mode. This causes the wormhole partition function to vanish.

In section \ref{sectwo} of this paper, we generalize \cite{Iliesiu:2021are} to the case of any supersymmetric partition function. The setup is as follows. Let $Z(X)$ be the partition function obtained by putting the boundary theory on manifold $X$ (which might include an $R$-current background, Wilson lines etc.). We will assume that $X$ has a continuous symmetry $Q$, which could be bosonic or fermionic. Now, consider the product of partition functions $Z(X) Z(X')$. The manifold $X \sqcup X'$ has a copy of the symmetry $Q$, acting only on $X$. The key point is that a smooth wormhole connecting the two boundaries will always spontaneously break this ``one-sided'' symmetry. This follows from the fact that a Killing vector or Killing spinor that vanishes on one boundary has to vanish everywhere.

To have an example in mind, one can think about the case where $X = X' = S^1$. Rotating the left boundary without rotating the right would introduce a twist in the wormhole:
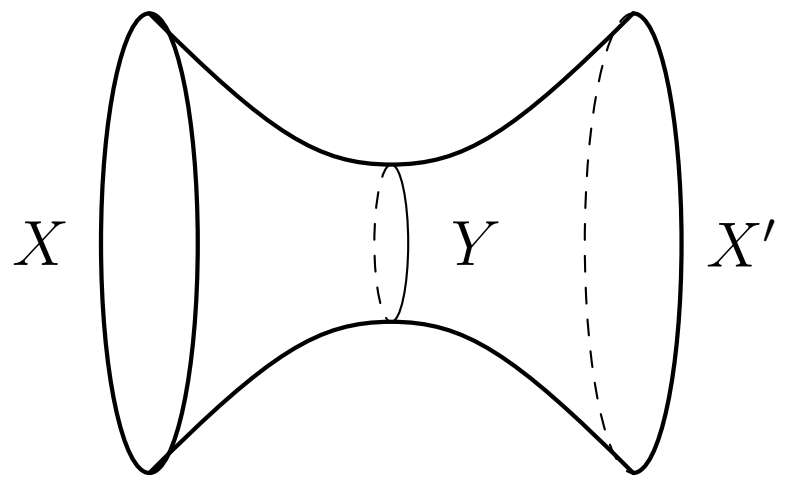
\begin{equation}
\begin{tikzpicture}[scale=.8, line cap=round,baseline=(current bounding box.center)]
  \draw[thick]
    (-2,1.9) .. controls (-1,0.9) and (-0.6,0.65) ..
    (0,0.65) .. controls (0.6,0.65) and (1,0.9) ..
    (2,1.9);

  \draw[thick]
    (-2,-1.9) .. controls (-1,-0.9) and (-0.6,-0.65) ..
    (0,-0.65) .. controls (0.6,-0.65) and (1,-0.9) ..
    (2,-1.9);

  \draw[thick] (-2,0) ellipse (0.4 and 1.9);

  \draw[dashed] (2,1.9) arc (90:270:0.4 and 1.9);
  \draw[thick]  (2,-1.9) arc (-90:90:0.4 and 1.9);

  \draw[dashed] (0,0.65) arc (90:270:0.14 and 0.65);
  \draw        (0,-0.65) arc (-90:90:0.14 and 0.65);

  \node at (-2.9,0) {$X$};
  \node at (2.9,0) {$X'$};
  \node at (.7,0) {$Y$};
\end{tikzpicture}
\end{equation}
So part of the symmetry is spontaneously broken.

This spontaneous breaking leads to zero modes. In the bosonic case, a time-translation zero mode of this type gives the ``ramp'' in the spectral form factor. In the fermionic case, the zero mode causes the wormhole to vanish. This can be argued precisely (for the case where the orbits of $Q^2$ have compact closure) using a localization theorem by Schwarz and Zaboronsky \cite{Schwarz:1995dg}. 

Our conclusion is that if $X$ preserves a supersymmetry, smooth wormholes will not contribute to $Z(X)Z(X')$. Apparently, supersymmetric partition functions are not ``erratic enough'' to earn wormhole contributions! Of course, this doesn't mean that supersymmetric partition functions must exactly factorize in any statistical ensemble. In section \ref{sec:variance_SYK_like_models} we study SYK-like examples where supersymmetric partition functions have a nonzero variance. There, the correlation $G_{LR}$ between two replicas is analogous to the wormhole in a gravitational theory. However, unlike a geometrical wormhole, a nonzero $G_{LR}$ can preserve all of the supersymmetry.

\section{Wormholes spontaneously break symmetries}\label{sectwo}

We will work in the setting of AdS/CFT. As in the introduction, let $Z(X)$ be a partition function in the boundary theory that preserves a symmetry $Q$. For the moment, we will allow $Q$ to be either bosonic or fermionic, but our primary interest is in the case where $Q$ is fermionic and the orbits of $Q^2$ have compact closure.\footnote{Examples include supersymmetric indices \cite{Romelsberger:2005eg,Kinney:2005ej,Assel:2014paa,Choi:2018hmj,Benini:2018ywd,Cabo-Bizet:2018ehj,Kim:2009wb}, topologically twisted indices \cite{Benini:2015noa,Benini:2015eyy,Benini:2016hjo,Closset:2016arn,Honda:2015yha,Hosseini:2016cyf}, certain sphere partition functions with or without Wilson line insertions \cite{Pestun:2007rz,Hama:2012bg,Bobev:2013cja,Kapustin:2009kz,Jafferis:2010un,Hama:2010av,Drukker:2010nc,Marino:2011eh,Freedman:2013oja,Hama:2011ea,Martelli:2011fu,Bobev:2022eus,Kallen:2012va,Jafferis:2012iv}, as well as supersymmetric partition functions on more complicated boundary manifolds \cite{Closset:2017zgf,Closset:2018ghr,Toldo:2017qsh,Crichigno:2018adf}.} In the holographic dual, this partition function corresponds to a boundary condition on the bulk manifold $Y$, namely that $\partial Y$ should have a component conformal to $X$. The fact that $X$ preserves the symmetry $Q$ means that this boundary condition preserves an asymptotic symmetry $Q$ of the bulk theory.

The boundary conditions to compute $Z(X)Z(X')$ also have a symmetry $Q$ that acts in the familiar way on $X$ but acts trivially on $X'$. We will show that any wormhole $Y$ connecting the two boundaries must spontaneously break this $Q$.

\subsection{Bosonic symmetries}
First consider the case of a bosonic symmetry. Then $Q$ corresponds to a (conformal) Killing vector $v$ of the boundary component $X$. Of course, we can regard this as a vector defined on the full boundary $X\sqcup X'$ that happens to vanish on $X'$. This symmetry is preserved by a bulk filling $Y$ if there is a bulk Killing vector $V$ that approaches $v$ at the boundary. To describe this more precisely, choose coordinates near a boundary $z\to 0$ so that
\begin{equation}
\d s^2 = \frac{\d z^2 + h_{ij} \d x^i \d x^j}{z^2} + \dots
\end{equation}
The bulk Killing vector equation implies that $V^z$ should vanish as $z\to 0$, and $V^i$ should approach a conformal Killing vector for the boundary metric $h$. To preserve $Q$ in the bulk, we would need $Y$ to have a Killing vector $V$ that approaches $v$ on the $X$ boundary component, and approaches zero on the $X'$ boundary component. However, this is impossible, because a Killing vector that vanishes on {\it any} boundary component has to vanish everywhere. This follows from the fact that if a Killing vector and its first derivatives vanish at any point on a connected manifold, then it is zero everywhere \cite[appendix C]{wald1984general}.

The conclusion is that a connected wormhole must spontaneously break any bosonic symmetry $Q$ that acts on only one of the boundaries.

\subsection{Fermionic symmetries}
For a fermionic symmetry, the logic is similar. Now  $Q$ is associated to a conformal Killing spinor $\epsilon$ for the manifold $X \sqcup X'$, with the property that $\epsilon$ vanishes on $X'$. For the wormhole $Y$ to preserve this symmetry, we would need $\epsilon$ to extend to a bulk Killing spinor $\mathcal{E}$. Roughly speaking, this Killing spinor would need to vanish near the $X'$ boundary but approach $\epsilon$ near the $X$ boundary. This is impossible because a Killing spinor that vanishes anywhere on a connected manifold vanishes everywhere.

To make this more precise, we will review what happens to a bulk Killing spinor near an AdS boundary, following \cite{Klare:2012gn}. Near $z\to 0$, the gravitino Killing spinor equation has the form 
\begin{equation}\label{kse}
0 = \left(\mathcal D_\mu+\frac{1}{2}\Gamma_\mu C\right)
    \mathcal E,
\end{equation}
where the matrix $C$ acts on the spinor and possibly internal supersymmetry indices. In the simplest case, $C=1$. It is convenient to work in radial gauge $A_z = 0$ for any gauge fields. Let $\gamma_i=e_i{}^a\Gamma_a$ denote the boundary gamma matrices, so that the bulk coordinate gamma matrices are $\Gamma_z=z^{-1}\Gamma_{\hat z}$ and $\Gamma_i=z^{-1}\gamma_i$. Then for $z\to 0$, (\ref{kse}) becomes
\begin{align}\label{kse2}
0 =
\left[z\partial_z+\frac{1}{2}\Gamma_{\hat z}C\right]\mathcal E, \hspace{20pt}
0 =
\left[zD_i+
\frac{\gamma_i}{2}\left(C-\Gamma_{\hat z}\right)
\right]\mathcal E.
\end{align}
Here $D_i$ denotes the spin and gauge covariant derivative for the boundary manifold, together with any additional terms arising from possible asymptotic background fields. The $-\Gamma_{\hat z}$ term in the second line came from the AdS spin connection. Near the boundary, we can make an ansatz
\begin{equation}
    \mathcal{E} = z^p \mathcal{E}_0(x) + O(z^{p+1}).
\end{equation}
Imposing (\ref{kse2}) leads to the following conclusions. First, $p = -1/2$. Second, $\Gamma_{\hat z} C\mathcal{E}_0(x) = \mathcal{E}_0(x)$, which is a spinor version of the statement that the ``normal component'' vanishes. Finally, $\mathcal{E}_0(x)$ satisfies the boundary conformal Killing spinor equation. The combination of these facts means that $\mathcal{E}_0(x)$ is the bulk spinor image of a nonzero boundary conformal Killing spinor $\epsilon(x)$.

So we learn that if a bulk Killing spinor is nonzero, it must approach a nonzero boundary conformal Killing spinor. It follows that a boundary conformal Killing spinor that vanishes on a component $X'\subset \partial Y$ cannot be extended to a Killing spinor of a connected $Y$. Therefore, as in the bosonic case, a connected wormhole must spontaneously break any fermionic symmetry $Q$ that acts only on one of the boundaries.

\subsection{A broken fermionic symmetry makes the wormhole vanish}
    We will now explain why the spontaneous breaking of a fermionic $Q$ implies that the associated wormholes vanish. The details of this argument have nothing to do with wormholes, they just rely on the fact that the integral over nontrivial orbits of a supersymmetry $Q$ vanish. A familiar manifestation of this is that if a saddle point spontaneously breaks a fermionic symmetry, then there is a fermionic zero mode, for which the integral vanishes. More generally, the fact that integrals over nontrivial orbits of $Q$ vanish underlies the mathematics of supersymmetric localization, where an integral can be localized to points where $Q$ vanishes and the orbits become trivial. We will now briefly review a proof of this statement given in \cite{Schwarz:1995dg}.

    Let $\Phi = \{B^i,\Psi^\alpha\}$ be bosonic and fermionic coordinates for a supermanifold, and suppose we have an integral invariant under a supersymmetry $Q$ that acts as
    \be 
  Q B^i = O(\Psi) , 
    \qquad 
    Q \Psi^\alpha = K^\alpha ( B) + O(\Psi^2) ,
    \label{eq:susy_transformation_general}
    \ee
    where $K^\alpha(B)$ is everywhere nonzero, meaning that there are no supersymmetric points in the integration space. We will also assume that $Q^2$ either acts trivially or generates a compact bosonic symmetry.    Choosing any positive metric on fermionic fields $g_{\alpha \beta}(\Phi)$ that is invariant under $Q^2$, we can define a Grassmann-odd function
    \be 
    \sigma (\Phi) \equiv g_{\alpha \beta}(\Phi) K^\alpha (B)  \Psi^\beta. \label{eq:sigma_definition}
    \ee
    This has the property that $Q\sigma$ does not vanish, because its ``number part'' (obtained by setting to zero all fermionic variables) is $g_{\alpha\beta}K^\alpha K^\beta$, which does not vanish by assumption. We can therefore define
    \be 
    \beta \equiv \frac{\sigma}{Q\sigma},
    \label{eq:function_beta}
    \ee
    which satisfies $Q\beta = 1$. Now the idea is that our integral can be written as
    \begin{equation}
\int \mu(\Phi) = \int Q(\beta \mu(\Phi)), 
\label{eq:Q_exact_answer}
    \end{equation}
    which vanishes provided that boundary contributions can be neglected. A similar conclusion applies to each orbit. If $b(B)$ is a bump function that selects a particular orbit, then a supersymmetrization is $\chi(\Phi) = Q(\beta b)$ and we can replace $\mu(\Phi)\to\chi(\Phi)\mu(\Phi)$ in the above.

    To prove the vanishing of wormholes, we apply this formalism to the case where $\Phi$ are the bulk fields on the wormhole topology. The symmetry breaking argument above implies that $K^\alpha(B)$ is nonzero. 
    
    We would now like to make a few comments about technical assumptions made in our argument. First, we assumed that the orbits of $Q^2$ are compact (or trivial). This was used in a subtle way in the above argument, by making it possible to choose a $Q^2$-invariant metric $g_{\alpha\beta}$ by averaging any positive metric over the compact orbits. As pointed out in \cite{Schwarz:1995dg} sometimes this is unnecessary.\footnote{Indeed, in the recent \cite{Maxfield:2026ynw}, the orbits of $Q^2$ are noncompact, but an explicit invariant metric is identified and localization works.} Second, we assumed that the integration space is compact, or at least that the relevant boundary terms vanish. This leaves open the possibility of a contribution from some degenerating limit of a wormhole (this is why we included the qualifier ``smooth wormhole'' in some places above). Typically, such limits are sensitive to the UV completion and it seems hard to say much about them, but see the Discussion for some comments about the case in JT gravity. Finally, we assumed the action of asymptotic boundary supersymmetry can be consistently defined in the bulk as an action on the properly extended gauge-fixed configuration space. The formalism to construct such an extension has been developed in \cite{deWit:2018dix} (see also \cite{Jeon:2018kec}). The proper application to our case would require extending the construction of \cite{deWit:2018dix} to a non-supersymmetric background. We expect such a procedure to be possible for any well-defined bulk supergravity theory.

\section{Variance of protected quantities in SYK-like models}
\label{sec:variance_SYK_like_models}

In the previous section, we showed that wormholes cannot contribute to supersymmetric quantities, because they necessarily break some of the supersymmetry. In this section, we will show that a type of ``non-geometric wormhole'' can contribute to the variance of supersymmetric quantities in SYK-like theories. (Here we are talking about the variance across the ensemble of couplings, not in the sense of $N$-averaging.) Concretely, this is a collective field configuration in which the correlation between the two replicas $G^{LR}$ is independent of the spacetime coordinates. Such a configuration can preserve all of the supersymmetry.

To study this, the simplest SYK example one might think of would be the Witten index in the ordinary $\mathcal{N} = 1$ or $\mathcal{N} = 2$ supersymmetric SYK models \cite{Fu:2016vas}. For these models, the index is exactly independent of the couplings and has no variance, so we have to study slightly more complicated cases. We will look at two cases. The first sounds more complicated but is actually simple: the $S^2$ partition function in the 2d SYK-like field theories with $(2,2)$ supersymmetry \cite{Chang:2023gow}. The second sounds simpler but is more complicated: the index in the $\mathcal{N} = 2$ SYK-like quantum mechanics models involving both bosons and fermions, studied in \cite{Biggs:2023mfn, Halder:2025hhi}.\footnote{In both cases, the variance in the supersymmetric partition function will come from fluctuations in $G_{LR}$ around zero (similar effects were also studied in the context of double-scaled SYK in \cite{Berkooz:2020fvm}). In other words, it does not come from a classical solution. It would be interesting to find or rule out examples where the variance comes from a nontrivial classical solution for $G_{LR}$.}

Before going into the details, we will give a high-level summary that applies to both calculations. In principle, to compute the ensemble average of the square of the supersymmetric partition function, we have to do a collective field path integral over superfields $G^{ab}(X,X')$, $\Sigma^{ab}(X,X')$, where $a,b \in \{L,R\}$ are replica indices and $X,X'$ are superspace coordinates. However, for the supersymmetric computations, the integral can be simplified dramatically, to an integral over supersymmetric configurations, parametrized by an ordinary bosonic matrix $G^{ab}$, in other words $G^{LL},G^{RR}$ and $G^{LR}$. In this integral, the variable $G^{LR}$ represents a type of non-geometric wormhole that preserves all of the supersymmetry acting on both the $L$ and $R$ systems.

\subsection{The \texorpdfstring{$S^2$}{S2} partition function in \texorpdfstring{$\mathcal{N} = (2,2)$}{N = (2,2)} SYK-like theories} \label{sub:S2}
We will start with the supersymmetric $S^2$ partition function for $\mathcal{N} = (2,2)$ theories in the $SU(2|1)_B$ background \cite{Doroud:2013pka,Gerchkovitz:2014gta,Ishtiaque:2017trm}. This was studied in \cite{Chang:2023gow} for a class of SYK-like theories that includes the $\mathcal{N} = 2$ versions \cite{Bulycheva:2018qcp} of the models studied in \cite{Murugan:2017eto}. These models are built out of $N$ chiral superfields $\Phi^i$ and $N$ anti-chiral superfields $\tilde\Phi^i$, with Lagrangian
\begin{equation}
\mathcal{L} = \mathcal{L}_{\rm kin} + \mathcal{L}_{W}.
\end{equation}
The kinetic and superpotential terms are
\begin{align}
\mathcal{L}_{\rm kin} = \int \d^2\theta \d^2\tilde\theta \tilde\Phi_i\Phi_i, \hspace{20pt} \mathcal{L}_{\rm W} = \i\int \d^2\theta W(\Phi_1,\dots,\Phi_N)\Big|_{\tilde\theta = \overline{\tilde\theta} = 0} + \rm{h.c.}
\end{align}
The supersymmetric configurations on $S^2$ (with appropriate background fields) are those in which the bottom $\phi^i$ and top $F^i$ components of $\Phi^i$ are constant and proportional to each other, and the $S^2$ partition function localizes to an integral over such configurations \cite{Gomis:2012wy}:\footnote{Gomis and Lee write the formula for twisted chiral multiplets. Exchanging chiral and twisted chiral multiplets gives the $SU(2|1)_B$ formula used in \cite{Chang:2023gow}; see also \cite{Ishtiaque:2017trm}.}
\begin{equation}
Z_{S^2} = \int_{\tilde\phi^i=(\phi^i)^*} \prod_{i=1}^N\d\phi^i\d\tilde\phi^i\,\exp\left[-4\pi\i W(\phi^1,\dots,\phi^N) -4\pi\i W^*(\tilde\phi^1,\dots,\tilde\phi^N)\right].
\end{equation}
More precisely, this formula is correct up to an ambiguous multiplicative constant that can be taken to be independent of the superpotential $W$.

For the SYK-like class we take
\begin{equation}
W = C_{i_1\dots i_p} \phi_{i_1}\dots \phi_{i_p}
\end{equation}
where all of the components of $C$ are identical circular complex random Gaussians with mean zero. We will use an overline bar to denote the average over these couplings. Then
\begin{equation}\label{Zaply}
\begin{aligned}
\overline{Z_{S^2}^k}
&=\int
\d^2\phi_i^a
\exp\left[-(\text{const.})N\sum_{a,b=1}^k(G^{ab})^p\right],\hspace{20pt} G^{ab}\equiv\frac1N\sum_{i=1}^N\phi_i^a(\phi_i^b)^*.
\end{aligned}
\end{equation}
The constant depends on the normalization of the distribution for the $C$ coefficients, and we will just take it to be one. To evaluate this integral, it is convenient to change variables from the $\phi_i^a$ to $G^{ab}$. To do this, one can insert $\delta(G^{ab} - \tfrac{1}{N}\phi_i^a\phi_i^{b*})$ and then integrate out the $\phi_i^a$. If these delta functions are implemented using integrals over Lagrange multiplier variables $\Sigma^{ab}$, the $\phi_i^a$ integrals are easy. For the present case, the $\Sigma^{ab}$ variable can also be integrated out using the Harish-Chandra-Itzykson-Zuber integral.\footnote{We learned this from AI \cite{AI}, and it was especially useful in the supermatrix version discussed later.} The result is significantly simpler than the intermediate steps, and is just (for $N > k$)
\begin{equation}\label{HCIZ}
\int_{\mathbb C^{kN}}\prod_{a=1}^k\prod_{i=1}^N\d^2\phi_i^a\,f(G)
=\frac{N^{Nk}\pi^{Nk-\frac{k(k-1)}{2}}}{\prod_{j=1}^k\Gamma(N-j+1)}
\int_{G > 0}\d G\,(\det G)^{N-k}f(G),
\end{equation}
where $\d^2\phi=\d\operatorname{Re}\phi\,\d\operatorname{Im}\phi$ and
\begin{equation}
\d G=\prod_{a=1}^k\d G^{aa}\prod_{a<b}\d\operatorname{Re}G^{ab}\,\d\operatorname{Im}G^{ab}.
\end{equation}
Applying this formula to (\ref{Zaply}) for $k = 1$, we get
\begin{equation}
\overline{Z_{S^2}} = \frac{(\pi N)^N}{\Gamma(N)}\int_0^\infty \d G G^{N-1}e^{-N G^p} = \frac{\pi^N N^{(1-\frac{1}{p})N }\Gamma(\frac{N}{p})}{p\Gamma(N)}.
\end{equation}
For the case $k = 2$, we get
\begin{equation}\label{sphere-two-replica-collective}
\begin{aligned}
\overline{Z_{S^2}^2}
={}&\frac{N^{2N}\pi^{2N-1}}{\Gamma(N)\Gamma(N-1)}
\int_0^\infty\d G^{LL}\int_0^\infty\d G^{RR}
\int_{|G^{LR}|^2<G^{LL}G^{RR}}\d^2G^{LR}\\
&\times\bigl(G^{LL}G^{RR}-|G^{LR}|^2\bigr)^{N-2}
\exp\left[-N\bigl((G^{LL})^p+(G^{RR})^p
+2\operatorname{Re}((G^{LR})^p)\bigr)\right].
\end{aligned}
\end{equation}
It is possible to evaluate this integral exactly \cite{AI}, but we will just study it for large $N$. In that limit, the $(\det G)^{N-2}$ factor in the integrand prefers $|G^{LR}| \ll G^{LL},G^{RR}$, which means that the $N \text{Re}((G^{LR})^p)$ term in the action can be considered a small perturbation. Expanding in this term, at zeroth other we get the factorized answer. At first order, we get zero, so the second-order term gives the leading contribution to the variance. For large $N$, this can be computed by approximating the $G^{LR}$ dependence of $(\det G)^{N-2}$ as Gaussian:
\begin{equation}
(\det G)^{N} \approx (G^{LL}G^{RR})^N e^{-N \frac{|G^{LR}|^2}{G^{LL}G^{RR}}} \approx (G^{LL}G^{RR})^N e^{-Np^{\frac{2}{p}}|G^{LR}|^2}
\end{equation}
where in the second equality we used the saddle point value of $G^{LL} = G^{RR} = (1/p)^{1/p}$.
Computing the expectation value of $\bigl(\text{Re}((G^{LR})^p)\bigr)^2$ in this distribution, one finds (for $ p>2$)
\begin{equation}
\frac{\text{var}(Z_{S^2})}{\overline{Z_{S^2}}^2} \approx \frac{\Gamma(1+p)}{p^2 N^{p-2}}.
\end{equation}
This leading answer for the variance can be explained entirely from fluctuations in the overall scale of the couplings $C$.

\subsection{The index in \texorpdfstring{$\mathcal{N} = 2$}{N=2} superfield SYK}
\label{sub:SYK}

\subsubsection{Average of the index}
\paragraph{The index from an integral over zero modes}
In quantum mechanics, the Witten index can be obtained as an integral over only the zero modes of the fields. The reason for this is that the index is independent of the inverse temperature $\beta$, and if we take $\beta$ to be very small, then for all of the modes except the zero modes, the kinetic terms will dominate over the potential terms and the integral over those variables becomes Gaussian. What remains is the integral over the zero modes:
\begin{equation}\label{zindex}
Z_{\text{index}} = \int \prod_{i=1}^N \frac{\d\phi^i \d F^i (-\i\d\psi_2^i\d\psi_1^i)}{2\pi} \exp\left[-\frac{F^iF^i}{2} - \i F^i \partial_{\phi_i}W + \i \psi_1^i \psi_2^j\partial_{\phi_i}\partial_{\phi_j}W\right].
\end{equation}
In this expression, there are a few technical points worth mentioning. First, we snuck in a factor of $1/(2\pi)$ in the measure. One can check that this is the result of integrating out all of the nonzero modes, starting with an arbitrary ultralocal measure. Second, we snuck in a factor of $(-\i)$ in the measure. This is in order to make the path integral agree with the Hilbert space definition of the index. In other words, for a pair of Majorana fermions the Hilbert space trace with $(-1)^F$ and the periodic path integral differ by a factor of $\i$. We inserted a factor here in order to make the path integral answer agree with the Hilbert space trace answer. Finally, we assumed that $W$ is homogeneous and we rescaled the fields  $\phi\to \lambda \phi$ and $\psi_i\to \lambda^{1/2}\psi_i$ in order to remove a $\beta$-dependent factor multiplying the $W$ terms in the action. (This rescaling doesn't change the measure.)

For a warmup example, consider the case $N = 1$ with $W = \frac{C}{2} \phi^2$. Then
\begin{equation}
Z_{\text{index}} = \int \frac{\d\phi\d F(-\i\d\psi_2\d\psi_1)}{2\pi}\exp\left[- \frac{F^2}{2} - \i C \phi F + \i C \psi_1\psi_2 \right] = \text{sgn}(C).
\end{equation}
For this example, if we average $C$ over a Gaussian distribution, the mean value of the index will be zero but the variance will be nonzero.

For the SYK application, we instead set
\begin{equation}
W = C_{i_1\dots i_p} \phi^{i_1}\dots \phi^{i_p}
\end{equation}
and then average over the tensor $C$. The averaging is simplest to define by treating all components as independent identically distributed real Gaussian variables, without imposing a symmetry constraint. Of course, only the symmetric part will survive. 

\paragraph{Collective field representation}
At this point it is useful to introduce some notation. After averaging $C_{i_1\dots i_p}$ over this distribution, the integrand will depend only on the bilinear $O(N)$ invariants 
\begin{equation}
G = \left(\begin{array}{cc|cc} G_{\phi\phi} & G_{\phi F} & G_{\phi 1} & G_{\phi 2} \\ G_{\phi F} & G_{FF} & G_{F1} & G_{F2} \\
\hline G_{\phi 1} & G_{F 1} & 0 & G_{12} \\ G_{\phi 2} & G_{F2} & -G_{12} & 0\end{array}\right)
\end{equation}
where the $G_{AB}$ are bilinears in the fundamental fields, averaged over $N$. For example, 
\begin{equation}\label{Gexampls}
G_{\phi\phi}=\frac{1}{N}\sum_{i=1}^N \phi^i\phi^i\hspace{20pt} \hspace{20pt} G_{F 1} = \frac{1}{N}\sum_{i=1}^N F^i\psi_1^i
\end{equation}
and similarly for the other components. It will also be useful to define a superfield bilocal
\begin{equation}
G(\theta,\theta')=\frac{1}{N}\sum_{i=1}^N\Phi^i(\theta)\Phi^i(\theta')
\end{equation}
where the fundamental superfield is
\begin{equation}
\Phi^i(\theta) = \phi^i + \i(\theta_1\psi_2^i - \theta_2\psi_1^i) - \theta_1\theta_2 F^i.
\end{equation}
Explicitly,
\begin{equation}
\begin{aligned}
G(\theta,\theta')={}&G_{\phi\phi}
+\i(\theta_1+\theta'_1)G_{\phi 2}
-\i(\theta_2+\theta'_2)G_{\phi 1}
-(\theta_1\theta_2+\theta'_1\theta'_2)G_{\phi F}
+(\theta_1\theta'_2-\theta_2\theta'_1)G_{12}\\
&+\i\theta_1\theta'_1(\theta_2-\theta'_2)G_{F2}
+\i\theta_2\theta'_2(\theta_1-\theta'_1)G_{F1}
+\theta_1\theta_2\theta'_1\theta'_2G_{FF}.
\end{aligned}
\end{equation}

Using this notation, we can compute the average of the action in (\ref{zindex}) by doing the Gaussian integral over the $C$ couplings, which we denote with an overline:
\begin{align}
\overline{\exp\left[-\i F^i \partial_{\phi_i}W + \i \psi_1^i \psi_2^j\partial_{\phi_i}\partial_{\phi_j}W\right]} &= \overline{\exp\left[\i \int \d\theta_2\d\theta_1 C_{i_1\dots i_p}\Phi^{i_1}(\theta)\dots \Phi^{i_p}(\theta)\right]}\\
&=\exp\left[-(\text{const})N \int \d\theta_2'\d\theta_1'\d\theta_2\d\theta_1 G(\theta,\theta')^p\right]
\end{align}
The constant depends on the normalization of the Gaussian measure for the $C$ couplings. It can be removed by rescaling the fields, and in what follows we will just set it equal to $1/p$. In terms of the components $G_{AB}$, the explicit formula for this superspace integral is
\begin{align}
\frac{1}{p}\int \d\theta_2'\d\theta_1'\d\theta_2\d\theta_1\,G(\theta,\theta')^p
={}&G_{FF}G_{\phi\phi}^{p-1}+(p-1)
\bigl(G_{\phi F}^2-G_{12}^2
-2G_{F1}G_{\phi 2}+2G_{F2}G_{\phi 1}\bigr)G_{\phi\phi}^{p-2}\notag\\
&-2(p-1)(p-2)
(G_{\phi F}+G_{12})G_{\phi 1}G_{\phi 2}G_{\phi\phi}^{p-3}.
\end{align}

As a next step, we can integrate out the fundamental fields holding fixed the bilinears, e.g.~(\ref{Gexampls}). As in the simpler case (\ref{HCIZ}) discussed above, it was pointed out to us by AI \cite{AI} that this procedure was already very conveniently ``precomputed'' in \cite{Sommers:2007}, which directly give the resulting measure for the $G$ supermatrix. The result is that the average over couplings $C$ of the index is
\begin{equation}\label{indexavg}
\overline{Z_{\text{index}}}
=\frac{(-\i)^N}{4\pi}\int \d G\,
(\operatorname{sdet}G)^{(N-1)/2}
\exp\left[-\frac{N}{2}G_{FF}
-\frac{N}{p}\int\d\theta'_2\d\theta'_1\d\theta_2\d\theta_1\,G(\theta,\theta')^p\right],
\end{equation}
where
\begin{equation}
\int \d G\equiv
\int_{G_{\mathrm B}>0}\d G_{\phi\phi}\d G_{\phi F}\d G_{FF}
\int\d G_{F2}\d G_{F1}\d G_{\phi2}\d G_{\phi1}
\oint_{0}\frac{\d G_{12}}{2\pi\i}.
\end{equation}
The matrix $G_B$ is defined as
\begin{equation}
G_{\mathrm B}=\begin{pmatrix}G_{\phi\phi}&G_{\phi F}\\G_{\phi F}&G_{FF}\end{pmatrix}.
\end{equation}
To define the possibly fractional power of the superdeterminant, let
\begin{equation}
\widehat G_{12}=G_{12}
-\begin{pmatrix}G_{\phi1}&G_{F1}\end{pmatrix}
G_{\mathrm B}^{-1}
\begin{pmatrix}G_{\phi2}\\G_{F2}\end{pmatrix}.
\end{equation}
Then
\begin{equation}
\operatorname{sdet}G=\frac{\det G_{\mathrm B}}{\widehat G_{12}^{\,2}},
\qquad
(\operatorname{sdet}G)^{(N-1)/2}
\equiv\frac{(\det G_{\mathrm B})^{(N-1)/2}}{\widehat G_{12}^{\,N-1}}.
\end{equation}

\paragraph{Semiclassical discussion}
The final step is to do the integral over the $G$ supermatrix. Let's first comment on the saddle point approximation to this, previously studied in section 3.1 of \cite{Halder:2025hhi}. We can make an ansatz where all of the Grassmann-odd components of $G$ are zero, and $G_{\phi F}$ is also zero. Then the action reduces to
\begin{equation}
-\frac{I}{N} = \frac{1}{2}\log \frac{G_{\phi\phi}G_{FF}}{G_{12}^2} - \frac{1}{2}G_{FF} - G_{FF}G_{\phi\phi}^{p-1} + (p-1)G_{12}^2G_{\phi\phi}^{p-2}.
\end{equation}
Imposing the equations of motion for $G_{FF}$ and $G_{12}$, we get
\begin{equation}\label{GFFG12}
G_{FF} = \frac{1}{1 + 2 G_{\phi\phi}^{p-1}}, \hspace{20pt} G_{12}^2 = \frac{1}{2(p-1)G_{\phi\phi}^{p-2}}
\end{equation}
and the resulting action, as a function of $G_{\phi\phi}$, is
\begin{equation}\label{actionGphiphi}
-\frac{I}{N} = \frac{1}{2}\log \frac{2(p-1) G_{\phi\phi}^{p-1}}{1 + 2 G_{\phi\phi}^{p-1}}.
\end{equation}
There is no saddle point of this action for finite $G_{\phi\phi}$, but there is a formal saddle point at infinity, with on-shell action $-\frac{I}{N} = \frac{1}{2}\log(p-1)$. This predicts that the expectation value of the index should be proportional to $(p-1)^{N/2}$ \cite{Biggs:2023mfn,Halder:2025hhi}. In fact there are actually two such formal saddles, with opposite signs for $G_{12}$, giving contributions proportional to $\i^N  + (-\i)^N$.

To make sense of this formal saddle point at infinity, one would like to include the measure for $G_{\phi\phi}$ that arises from doing the integral over small fluctuations in the other components of the superfield, and then integrate over $G_{\phi\phi}$. In fact, it turns out it to be possible to do the integral over all of the other components exactly \cite{AI}, as we explain in the next section.

\paragraph{Exact computation}
As a first step, one can integrate out the Grassmann-odd entries by just extracting the coefficient of $G_{\phi1}G_{\phi2}G_{F1}G_{F2}$ in the integrand. A useful formula is that the shifted $\widehat G_{12}$ obeys
\begin{equation}
(G_{12}-\widehat G_{12})^2
=\frac{2G_{\phi1}G_{\phi2}G_{F1}G_{F2}}{\det G_{\mathrm B}},
\end{equation}
and therefore
\begin{equation}
\frac{1}{\widehat G_{12}^{\,N-1}}
=\frac{1}{G_{12}^{N-1}}\left[
1+\frac{N-1}{G_{12}}(G_{12}-\widehat G_{12})
+\frac{N(N-1)}{\det G_{\mathrm B}\,G_{12}^2}
G_{\phi1}G_{\phi2}G_{F1}G_{F2}\right].
\end{equation}
After expanding the exponential out in powers of the Grassmanns, combining with this factor, and retaining the coefficient of $G_{\phi1}G_{\phi2}G_{F1}G_{F2}$, we find
\begin{equation}\label{index-after-odd-G}
\begin{aligned}
\overline{Z_{\text{index}}}
={}&\frac{(-\i)^N N}{4\pi}
\int_{G_{\mathrm B}>0}\d G_{\phi\phi}\d G_{\phi F}\d G_{FF}
(\det G_{\mathrm B})^{(N-3)/2}\\
&\times\oint_{0}\frac{\d G_{12}}{2\pi\i\,G_{12}^{N+1}}
\exp\left[-N\left(\frac12+G_{\phi\phi}^{p-1}\right)G_{FF}
-N(p-1)G_{\phi\phi}^{p-2}(G_{\phi F}^2-G_{12}^2)\right]\\
&\times\Bigl\{N-1+2(p-1)G_{\phi\phi}^{p-2}G_{12}^2
\left[(N-1)(p-2)-2N(p-1)G_{\phi\phi}^{p-2}\det G_{\mathrm B}\right]\Bigr\}.
\end{aligned}
\end{equation}
In this computation, we dropped a term that was odd in $G_{\phi F}$ and that therefore integrates to zero in a later step. The $G_{12}$ contour integral vanishes for odd $N$. For even $N$, it is
\begin{equation}\label{index-after-G12}
\begin{aligned}
\overline{Z_{\text{index}}}
={}&\frac{(-1)^{N/2}N(p-1)}{4\pi\,\Gamma(\frac{N}{2}+1)}
\int_{G_{\mathrm B}>0}\d G_{\phi\phi}\d G_{\phi F}\d G_{FF}\exp\left[-N\left(\frac12+G_{\phi\phi}^{p-1}\right)G_{FF}
-N(p-1)G_{\phi\phi}^{p-2}G_{\phi F}^2\right]\\
&\times(\det G_{\mathrm B})^{(N-3)/2}
\bigl[N(p-1)G_{\phi\phi}^{p-2}\bigr]^{N/2}
\left[N-1-2NG_{\phi\phi}^{p-2}\det G_{\mathrm B}\right].
\end{aligned}
\end{equation}
Only the three bosonic entries of $G_{\mathrm B}$ remain to be integrated. It is convenient to change variables from $G_{FF}$ to $\det G_{\rm B}$. The condition that $G_{\rm B}$ is a positive matrix implies that $G_{\phi\phi} > 0, \det G_{\rm B} > 0$. The integral over $\det G_{\rm B}$ is then a Gamma function-type integral, and the integral over $G_{\phi F}$ is Gaussian. The final result, in a form that is valid for both even and odd $N$, is
\begin{align}\label{indexfinal}
\overline{Z_{\text{index}}}
&=\left(\i^N + (-\i)^N\right)Z_{N,p}\\
Z_{N,p} &\equiv \frac{(p-1)\Gamma\bigl(\frac{N+1}{2}\bigr)}
{4\sqrt{\pi}\,\Gamma(\frac{N}{2})}\int_0^\infty\frac{\d G_{\phi\phi}}{G_{\phi\phi}\sqrt{\frac12+G_{\phi\phi}^{p-1}}
\sqrt{\frac12+pG_{\phi\phi}^{p-1}}}\,
\exp\left[\frac{N}{2}\log\frac{2(p-1)G_{\phi\phi}^{p-1}}{1+2G_{\phi\phi}^{p-1}}\right].
\end{align}
Note that the integrand contains the action (\ref{actionGphiphi}) as a function of $G_{\phi\phi}$, but now corrected by the measure obtained by properly integrating out all of the other variables. This measure makes the integral convergent. As an example, for $p = 3,N = 12$, one finds $\overline Z_{\rm index} = 1 + \frac{9\sqrt{3}}{2}$.

Note that the index equals the ``topological degree'' of $\nabla W$. For even $N$, our result agrees with the exact mean of this degree obtained by Khimshiashvili in Theorem 4 of \cite{Khimshiashvili:2004}.

\subsubsection{Variance of the index}
One can write an exact formula for $\overline{Z_{\rm index}^2}$ similar to (\ref{indexavg}). The difference is that now the $G$ supermatrix contains blocks corresponding to the $LL, LR, RR$ correlators, and the superdeterminant is over this larger space. The action includes the substitution
\begin{equation}
G(\theta,\theta')^p \to \sum_{a,b} G^{ab}(\theta,\theta')^p = G^{LL}(\theta,\theta')^p +  G^{RR}(\theta,\theta')^p +  G^{LR}(\theta,\theta')^p+G^{LR}(\theta',\theta)^p.
\end{equation}
We will compute the variance of the index for large $N$, in the same way as for the simpler computation in section \ref{sub:S2}. The main idea is to expand in the $G_{LR}^p$ interaction appearing in the action. For large $N$, the $G_{LR}$ variables will be small and approximately Gaussian distributed, with a covariance given by the statistics in the factorized theory where there is no $LR$ correlation:
\begin{align}\label{variance-LR-covariance}
\left\langle G^{LR}(\theta,\theta')G^{LR}(\chi,\chi')\right\rangle_{0}
&=\frac{1}{N^2} \sum_{i,j} \langle \Phi^{iL}(\theta)\Phi^{iR}(\theta')\Phi^{jL}(\chi)\Phi^{jR}(\chi')\rangle_0\\
&=\frac{1}{N^2} \sum_{i} \langle \Phi^{iL}(\theta)\Phi^{iL}(\chi)\rangle_0\langle\Phi^{iR}(\theta')\Phi^{iR}(\chi')\rangle_0\\
&=\frac1N G^{LL}(\theta,\chi)G^{RR}(\theta',\chi').
\end{align}
If life were simple, then for large $N$ we would just plug into the RHS the saddle point values of the diagonal $LL,RR$ fields. However, recall that for the index computation there is actually no saddle point for the $G^{LL},G^{RR}$ variables. There is the next best thing, though, which is a saddle point for all of the variables except for the $G_{\phi\phi}$ component, which has a sort of saddle point at infinity. Along this locus, some of the components of the superfield are zero, and
\begin{equation}
G^{aa}(\theta,\xi)
=G_{\phi\phi}^{aa}
+G_{12}^{aa}(\theta_1\xi_2-\theta_2\xi_1)
+G_{FF}^{aa}\theta_1\theta_2\xi_1\xi_2,
\qquad a=L,R,
\end{equation}
where $G_{FF}$ and $G_{12}$ are given in (\ref{GFFG12}) in terms of $G_{\phi\phi}$. It will be useful to know that
\begin{equation}
G_{FF}^{aa}(G_{\phi\phi}^{aa})^{p-1}
=\frac{(G_{\phi\phi}^{aa})^{p-1}}{1+2(G_{\phi\phi}^{aa})^{p-1}}
\longrightarrow\frac12,
\qquad
(G_{12}^{aa})^2(G_{\phi\phi}^{aa})^{p-2}=\frac1{2(p-1)}.
\end{equation}
An important detail is that for a given value of $G_{\phi\phi}$ there are actually two saddle points, distinguished by the sign of $G_{12}$. We will have to sum over these independently for the $LL$ and $RR$ sectors. This sum will lead to interesting dependence on the parity of $p,N$. For each of these choices of saddle, we can expand in the $(G^{LR})^p$ term in the action, and we would like to compute
\begin{equation}
\left\langle 1 - \frac{2N}{p}\int_{\theta,\theta'} G^{LR}(\theta,\theta')^p + \frac{(2N)^2}{2p^2}\int_{\theta,\theta'} G^{LR}(\theta,\theta')^p\int_{\chi,\chi'} G^{LR}(\chi,\chi')^p + \dots\right\rangle_0.
\end{equation}

\paragraph{First order term}
Let's study the first nontrivial term in this expansion. If $p$ is odd, then this term does not contribute. But if $p$ is even, then it can contribute, and we find
\begin{align}
\left\langle\int_{\theta,\theta'} G^{LR}(\theta,\theta')^p\right\rangle &= \frac{(p-1)!!}{N^{p/2}}\int \d\theta'_2\d\theta_1'\d\theta_2\d\theta_1(G_{\phi\phi} + 2\theta_1\theta_2G_{12}^{LL})^{\frac{p}{2}}(G_{\phi\phi} + 2\theta_1'\theta_2'G_{12}^{RR})^{\frac{p}{2}}\notag\\
&=\frac{p^2(p-1)!!}{N^{p/2}} G_{12}^{LL}G_{12}^{RR} G_{\phi\phi}^{p-2}\\
&= \frac{p^2(p-3)!!}{2N^{p/2}} \text{sgn}(G_{12}^{LL})\text{sgn}(G_{12}^{RR}).
\end{align}
Now let's consider the sum over the signs of $G_{12}^{aa}$. Naively, this gives zero. That is the right answer for even $N$. But for odd $N$, recall that the two saddles contribute with opposite signs (this is why the average of the index is zero), so they actually add together. We conclude that
\begin{equation}\label{noddpeven}
\text{var}(Z_{\rm index}) \approx \frac{4p(p-3)!!}{N^{p/2-1}}Z_{N,p}^2, \hspace{20pt} (N \gg 1 \text{ odd},\, p\text{ even} \ge 4).
\end{equation}

\paragraph{Second order term}
Within the second order term, we have to sum over a set of distinct Wick contractions. Suppose that $r$ of the $p$ fields at one vertex contract with fields at the other vertex. The rest contract within their own vertices, so $p-r$ must be even. The number of Wick contractions is
\begin{equation}
\binom pr^2 r!\bigl[(p-r-1)!!\bigr]^2,
\end{equation}
with $(-1)!!=1$. For a given Wick contraction, we have to compute the Grassmann integral
\begin{equation}\label{varvar}
\begin{aligned}
&\int\d^2\theta\,\d^2\xi\,
G^{aa}(\theta,\theta)^{\frac{p-r}{2}}
G^{aa}(\xi,\xi)^{\frac{p-r}{2}}
G^{aa}(\theta,\xi)^r\\
&\quad=rG_{FF}^{aa}(G_{\phi\phi}^{aa})^{p-1}
+\bigl[p^2-(2p-1)r\bigr](G_{12}^{aa})^2(G_{\phi\phi}^{aa})^{p-2}
\longrightarrow\frac{p(p-r)}{2(p-1)}.
\end{aligned}
\end{equation}
Note that this result is independent of the sign of $G_{12}^{aa}$. The sum over signs will then vanish for odd $N$, but will be nonzero for even $N$, giving
\begin{align}
\text{var}(Z_{\rm index}) &\approx 4\frac{(2N)^2}{2p^2}\frac{1}{N^{p}}
\sum_{\substack{0\leq r\leq p\\r=p\ ({\rm mod}\ 2)}}
\binom pr^2 r!\bigl[(p-r-1)!!\bigr]^2
\left[\frac{p(p-r)}{2(p-1)}\right]^2 Z_{N,p}^2\\
&=\frac{2p^2(2p-5)!!}{N^{p-2}}Z_{N,p}^2,
\qquad (N\gg 1\text{ even},\,p\geq3).\label{neven}
\end{align}

For $N$ odd and $p$ odd, we find that both the first and second order terms vanish, so
\begin{equation}
\text{var}(Z_{\rm index}) \approx 0,\hspace{20pt} (N\text{ odd},\,p \text{ odd}).
\end{equation}

\paragraph{Exact computation} Note that we have not actually written an exact formula for the two-replica answer as an integral over the supersymmetric locus $G^{ab}_{\phi\phi}$ generalizing (\ref{indexfinal}) in the one replica case. It is possible to do this \cite{AI}, but we will not report the full formula here.

\section{Discussion}

In this paper, we showed that wormholes connecting multiple asymptotic boundaries do not contribute to supersymmetric partition functions, because they spontaneously break part of the supersymmetry.\footnote{Some wormholes in the literature, such as \cite{Astesiano:2023iql}, preserve a linear combination of the $L$ and $R$ supersymmetries, but not e.g.~the $L$ supersymmetry by itself.} Of course, such wormholes could still capture interesting information about non-supersymmetric quantities containing additional insertions that could saturate the fermionic zero mode.

We focused on the case with asymptotically AdS boundaries, but we expect that the argument can be generalized to more general asymptotics. Indeed, our results have an antecedent in the case of asymptotically flat wormholes that have been argued to generate random EFT couplings \cite{Coleman:1988cy,Giddings:1988cx}. In supersymmetric theories, the type of couplings that are generated this way are ``D-term'' couplings of the schematic form
\begin{equation}
I \supset \alpha \int \d^d x  \d^\mathcal{N}\zeta \mathcal{O}(x,\zeta).
\end{equation}
Here, the integrals $\d^d x \d^\mathcal{N}\zeta$ are over orbits of the symmetry broken by each wormhole mouth. See explicit computations in \cite{Anglin:1992ym}. It was suggested in \cite{Park:1990ep,Martucci:2024trp} that ``F-term'' couplings could be generated in cases where the wormhole degenerates, and some of the fermion zero modes are lost. 

\begin{figure}[t]
\centering
\begin{tikzpicture}[
  scale=.8,
  line cap=round,
  baseline=(current bounding box.center)
]
  \draw[thick]
    (-4.75,1.9) .. controls (-3.55,0.4) and (-1.85,0) ..
    (-0.65,0);
  \draw[thick]
    (-4.75,-1.9) .. controls (-3.55,-0.4) and (-1.85,0) ..
    (-0.65,0);

  \draw[thick]
    (0.65,0) .. controls (1.85,0) and (3.55,0.4) ..
    (4.75,1.9);
  \draw[thick]
    (0.65,0) .. controls (1.85,0) and (3.55,-0.4) ..
    (4.75,-1.9);

  \draw[thick, decorate,
    decoration={snake, amplitude=1.5pt, segment length=6pt,
                pre length=1pt, post length=1pt}]
    (-0.48,0) -- (0.48,0);

  \draw[thick] (-4.75,0) ellipse (0.4 and 1.9);

  \draw[dashed] (4.75,1.9) arc (90:270:0.4 and 1.9);
  \draw[thick]  (4.75,-1.9) arc (-90:90:0.4 and 1.9);

  \node at (-5.65,0) {$X$};
  \node at (5.65,0) {$X'$};
\end{tikzpicture}
\caption{Correlated Ramond punctures would give a nonzero variance to the index in JT supergravity.}\label{cusps}
\end{figure}
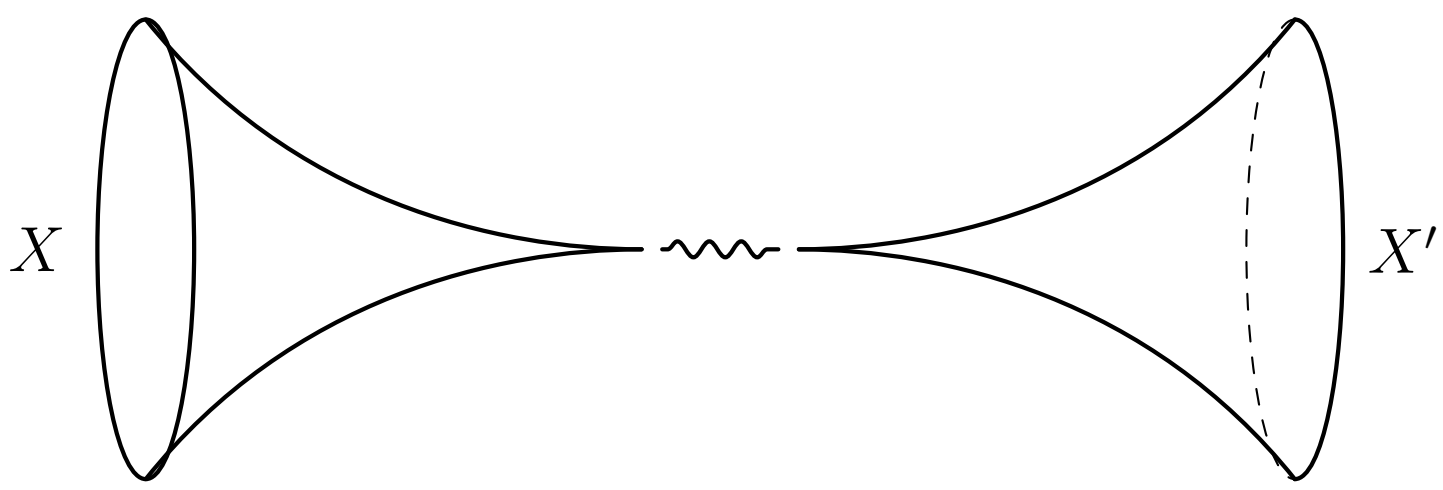

We can get some insight into this from JT supergravity, and in particular from the vanishing contribution of a cylinder wormhole (double trumpet) to the square of the Witten index \cite{Stanford:2019vob}. Let $b$ be the length of the geodesic ``waist'' of the wormhole. Then $b\to 0$ is a degenerating limit where the wormhole approaches two infinite cusps. The fermion zero mode exists for any $b$, and the integral over $b$ all the way to $b = 0$ vanishes exactly. However, there can be a nonzero contribution from a configuration with no fermion zero modes that is similar to the degenerating limit itself: two $b = 0$ infinite cusps with Ramond spin structures (Ramond punctures). This configuration factorizes into two disconnected manifolds, and of course there can also be contribution to a single copy of the Witten index from a single cusp. In the simplest treatment, there is a fixed fugacity for Ramond punctures, which becomes a fixed value of the Witten index. But it would also make sense to define a separate fugacity for pairs of cusps etc, leading in general to a nonzero variance for the index, see figure \ref{cusps}.\footnote{A similar situation can be arranged in 3d pure supergravity \cite{Stanford:2025llj}.}

Finally, we come back to our original motivation for this paper. The conjecture that wormholes are related to erratic behavior in the large $N$ limit \cite{Schlenker:2022dyo,Liu:2025ikq, Kudler-Flam:2026nzz,Liu:2026fnd}, together with the absence of wormhole contributions to supersymmetric quantities, suggests that supersymmetric quantities are not ``sufficiently erratic,'' or perhaps not erratic in the ``right'' way. It would be interesting to investigate this from the boundary theory side.\footnote{For example, for the case of the superconformal index in $\mathcal{N} = 4$ SYM, one might naively argue that the giant graviton expansion \cite{Gaiotto:2021xce,Imamura:2021ytr} implies a representation in terms of non-erratic terms of the form $e^{-a N}N^{-b}$. However, this series can diverge in certain regions of fugacity space, see \cite{Lee:2026zew} for recent discussion.}

\section*{Acknowledgements}

We are grateful to  Luca V. Iliesiu, Maciej Kolanowski, Ji Hoon Lee, Elisa Tabor, Gustavo Joaquin Turiaci, Cynthia Yan, and Zhenbin Yang. This work was supported in part by DOE grant DE-SC0026143 and by a grant from the Simons foundation (926198, DS). JB is supported by the Leinweber Foundation and the Leinweber Institute for Theoretical Physics at Stanford University.

\appendix

\section{Numerical computations of the index in superfield SYK}
\begin{figure}[H]
\centering
\begin{tikzpicture}[x=0.155cm,y=0.065cm]
 \draw[->] (-48,0) -- (48,0) node[right] {$Z_{\rm index}$};
 \draw[->] (-48,0) -- (-48,49);
 \foreach \x/\h in {-45/1,-35/6,-33/6,-31/4,-29/3,-27/14,-25/6,-23/11,-21/13,-19/20,-17/15,-15/24,-13/28,-11/29,-9/36,-7/37,-5/44,-3/44,-1/46,1/39,3/33,5/41,7/42,9/28,11/24,13/21,15/20,17/21,19/12,21/18,23/6,25/9,27/3,29/7,31/4,33/3,35/2,37/1,39/2,41/2,43/2,45/1} {
  \filldraw[fill=blue!45,draw=blue!70!black]
   (\x-0.75,0) rectangle (\x+0.75,\h);
 }
 \foreach \x in {-45,-35,-25,-15,-5,5,15,25,35,45} {
  \draw (\x,0) -- (\x,-0.18) node[below,font=\small] {\x};
 }
 \foreach \y in {0,10,20,30,40} {
  \draw (-48,\y) -- (-48.5,\y) node[left,font=\small] {\y};
 }
\end{tikzpicture}
\caption{Histogram of $Z_{\rm index}$ for $p=4$ and $N=9$, with $728$ samples.}\label{figodd}
\end{figure}
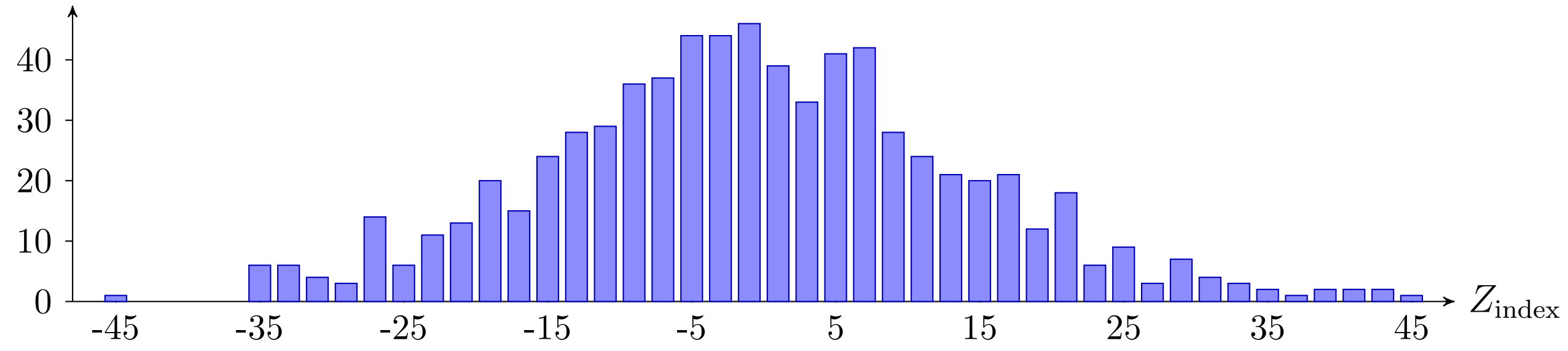

\begin{figure}[H]
\centering
\begin{tikzpicture}[x=0.105cm,y=0.055cm]
 \draw[->] (-20,0) -- (122,0) node[right] {$Z_{\rm index}$};
 \draw[->] (-20,0) -- (-20,34);
 \foreach \x/\h in {-16/1,-8/5,-6/4,-4/2,-2/8,0/10,2/15,4/12,6/15,8/17,10/32,12/22,14/23,16/21,18/25,20/16,22/28,24/29,26/23,28/27,30/27,32/18,34/22,36/24,38/31,40/21,42/26,44/20,46/22,48/14,50/13,52/9,54/6,56/9,58/6,60/11,62/7,64/5,66/3,68/3,70/3,72/3,74/7,76/2,78/4,80/3,82/3,84/5,86/1,92/1,94/1,98/1,104/1,106/1,118/1} {
  \filldraw[fill=blue!45,draw=blue!70!black]
   (\x-0.75,0) rectangle (\x+0.75,\h);
 }
 \foreach \x in {-16,0,16,32,48,64,80,96,112} {
  \draw (\x,0) -- (\x,-0.08) node[below,font=\small] {\x};
 }
 \foreach \y in {0,5,10,15,20,25,30} {
  \draw (-20,\y) -- (-20.5,\y) node[left,font=\small] {\y};
 }
\end{tikzpicture}
\caption{Histogram of $Z_{\rm index}$ for $p=3$ and $N=16$, with 669 samples.}\label{figNeven}
\end{figure}
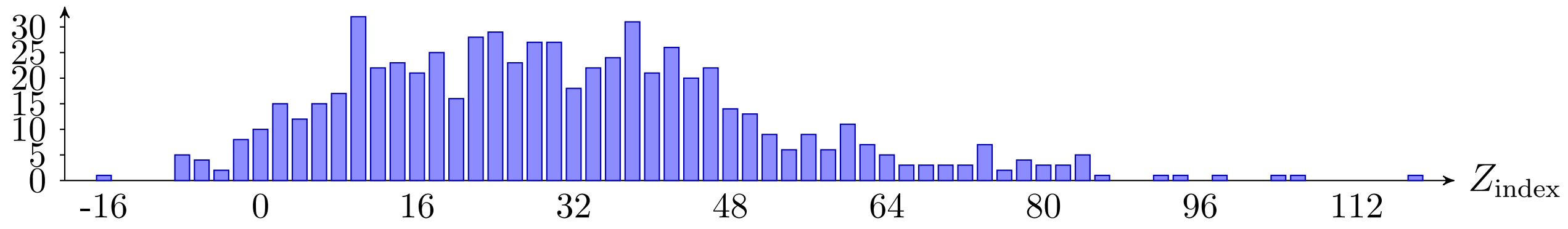

For some small values of $p,N$, we sampled the index numerically, by using the Julia package \texttt{HomotopyContinuation.jl} \cite{breiding2018homotopycontinuationjlpackagehomotopycontinuation} to find all of the critical points. The index is simply the sum over real critical points, weighted by the sign of the determinant of the Hessian. To resolve the degenerate critical point at the origin, we add an arbitrary linear term $W \to W + v^i \phi_i$. For the case of $N$ odd and $p$ odd, the index is exactly zero, configuration by configuration \cite{Biggs:2023mfn}. This is consistent with our findings above. For the case of $N$ odd and $p$ even, the mean of the index is zero, but the variance is nonzero, see figure \ref{figodd} for the case $p = 4,N = 9$. In that data, the empirical mean and variance are -0.6 and 211. The exact mean (\ref{indexfinal}) is 0, and the large $N$ answer for the variance (\ref{noddpeven}) is approximately 172.
For the case of even $N$, the mean and variance are nonzero, see figure \ref{figNeven} for the case $p = 3,N = 16$. From this data the empirical mean and variance are 30.9 and 432. The exact mean is 30.2284\dots, and the large $N$ approximation to the variance is approximately 257. This is uncomfortably far from the empirical variance, but apparently it is due to large $1/N$ corrections. It was reported by AI \cite{AI} that the exact answer for the variance is $446.71\dots$.

\bibliographystyle{utphys2}
{\small \bibliography{Biblio}{}}

\end{document}